\documentclass[11pt]{elsart}
\usepackage{setspace}
\usepackage{url}
\usepackage{bm}
\usepackage{cases}
\usepackage[dvips]{graphicx}

\begin{document}
\begin{frontmatter}

\title{Identifying influential spreaders in complex networks based on  gravity formula}
\author{Ling-ling Ma$^{a}$}
\author{Chuang Ma$^{a}$}
\author{Hai-Feng Zhang$^{a,b,c}$}\footnote{Corresponding author:
haifengzhang1978@gmail.com}
\author{Bing-Hong Wang$^{d}$}
\address
{$^{a}$ School of Mathematical Science, Anhui University, Hefei
230601, P. R. China }

\address{$^{b}$  Key Laboratory of Computer Network and Information Integration (Southeast University), Ministry of Education, Nanjing
211189, P. R. China }
\address {$^{c}$ Department of Communication Engineering, North University of China, Taiyuan, Shan'xi 030051, P. R. China}
\address{$^{d}$ Department of Modern Physics, University of Science and
Technology of China, Hefei 230026,  China}

\begin{abstract}

How to identify the influential spreaders in social networks is crucial for accelerating/hindering information diffusion, increasing product exposure, controlling diseases and rumors, and so on.  In this paper, by viewing the k-shell value of each node as its mass and the shortest path distance between two nodes as their distance, then inspired by the idea of the gravity formula, we propose a gravity centrality index to identify the influential spreaders in complex networks. The comparison between the gravity centrality index and some well-known centralities, such as degree centrality, betweenness centrality, closeness centrality, and k-shell centrality, and so forth, indicates that our method can effectively identify the influential spreaders in real networks as well as synthetic networks. We also use the classical Susceptible-Infected-Recovered (SIR) epidemic model to verify the good performance of our method.
\end{abstract}

\date{}
\end{frontmatter}

\section{Introduction} \label{sec:intro}

To effectively identify influential spreaders in social networks is of theoretical and practical significance~\cite{kitsak2010identification,basaras2013detecting,lu2011leaders,borge2012absence,klemm2012measure,wei2013identifying,liu2013ranking,chen2013path,ren2014iterative,zhong2015iterative,zhao2014identifying}, since it is crucial for developing efficient strategies to control epidemic spreading, accelerate information diffusion, promote new products, and so on. In view of this, many centrality indices have been proposed to address this problem, including degree centrality~\cite{bonacich1972factoring}, betweenness centrality~\cite{freeman1977set}, neighborhood centrality~\cite{maslov2002specificity} and closeness centrality~\cite{sabidussi1966centrality}, etc. In particular, Kitsak \emph{et~al}. proposed a \emph{k}-shell decomposition method to identify the most influential spreaders based on the assumption that nodes
in the same shell have similar influence and nodes in higher
shells are likely to infect more nodes. k-shell method is found to be better than the degree centrality index in many real networks~\cite{kitsak2010identification}. However, recent researches have demonstrated that the nodes within the same shell often have distinct influences, and this method may fail in some networks without core-like structure, e.g., Baras\'{a}si-Albert network~\cite{barabasi1999emergence}. Thus, after this, some methods were proposed to further improve the performance of the k-shell method. For example, Zeng \emph{et~al}. proposed a mixed degree decomposition method by incorporating the residual degree and the exhausted degree~\cite{zeng2013ranking}; Liu \emph{et~al} have demonstrated that the existence of the core-like groups can result in the invalidation of k-shell method~\cite{liu2014core}, and then they showed that the accuracy of k-shell method can be improved once the redundant links in networks are removed~\cite{liu2015improving}.  Chen \emph{et al}. designed a semi-local index by considering the next nearest neighborhood~\cite{chen2012identifying}; Lin \emph{et~al}. presented an improved ranking method by taking into account the shortest path distance between a target node and the node set with the highest k-core value~\cite{lin2014identifying}; Recently, Bae \emph{et~al}. defined a novel measure--coreness centrality index, which is given by summing all neighbors' k-shell values~\cite{bae2014identifying}.

In general, a node's influence is not only dependent on its nearest neighbors but also on the nodes who are not the nearest neighbors~\cite{zhang2011node,zhang2009seeding}, meanwhile, their interaction influence commonly decreases with their shortest path distance. If the k-shell value of each node is viewed as its mass, and the shortest path distance between two nodes is defined as their distance, then we can use the idea of gravity formula proposed by Isaac Newton to measure the influence of nodes. Inspired by these factors, in the work, we propose a new centrality index to measure the influence of nodes, which is called gravity centrality index. We apply the susceptible-infectious-recovered (SIR) spreading dynamics to evaluate the effectiveness of our proposed method, the experimental results indicate that gravity centrality index can better evaluate the influence of nodes than the ones generated by degree centrality, betweenness centrality, k-shell centrality, closeness centrality, and so on.

The layout of the paper is as follows: In Sec.~\ref{sec:method}, we first briefly review several typical centrality indices which are used to compare in this work, and the description of our method is presented. Then the experimental results are presented in Sec.~\ref{sec:main results}. Finally,
Conclusions and discussions are summarized in
Sec.~\ref{sec:discussion}.

\section{Method} \label{sec:method}
An undirected network is represented by $G=(N,M)$ with $N$ nodes and $M$ edges, and its structure can be described by an adjacent matrix $A=({a_{ij}})_{N\times N}$ where $a_{ij}=1$ if node $i$ is connected to node $j$, and $a_{ij}=0$ otherwise.

Here we briefly review the definitions of several centrality indices that will be discussed in this work.

The degree centrality (DC) of a node is defined as the number of nearest neighbors. The betweenness centrality (BC) of a node is defined as the fraction of all shortest paths travel through the node. The closeness centrality (CC) of a node is defined as the reciprocal of the sum of the lengths of the geodesic distance to every other node. The k-shell decomposition method (ks) is implemented by the following steps: Firstly, remove all nodes with degree one, and keep deleting the existing nodes until all nodes' degrees
are larger than one. All of these removed nodes are assigned 1-shell. Then \emph{recursively} remove the nodes with degree no larger than two (i.e., remove all nodes with degree two, and keep deleting the existing nodes until all nodes' degrees
are larger than two.) and include them to 2-shell. This procedure continues until all nodes
have been assigned to one of the shells~\cite{zeng2013ranking}.

To improve the exactness of k-shell method, the mixed degree decomposition (MDD) method was proposed by Zeng \emph{et~al.} ~\cite{zeng2013ranking}. The mixed degree $k_{m}(i)$ for a node $i$ is defined by considering the residual degree $k_r(i)$ and the exhausted degree $k_e(i)$ simultaneously, which is written as:
\begin{equation}\label{1}
k_{m}(i)=k_{r}(i)+\lambda \ast k_{e}(i).
\end{equation}
At each step of the MDD procedure, the nodes are removed according to the mixed degree, and the mixed degrees of remaining nodes are also updated. Where
 $\lambda$ is a tunable parameter between 0 and 1. As in Ref.~\cite{zeng2013ranking}, we take $\lambda=0.7$ in this work.

Recently, Baus \emph{et~al.} designed a ranking method--neighborhood coreness $C_{nc}$ by considering the degree and the coreness of a node simultaneously, the $C_{nc}(i)$ for a node $i$ is defined as~\cite{bae2014identifying}
\begin{equation}\label{2}
C_{nc}(i)=\sum _{j\in \Lambda_i} ks(j),
\end{equation}
where $\Lambda_i$ is the neighbor node set of node $i$. They further developed an extended neighborhood coreness $C_{nc+}$, which is described as:
\begin{equation}\label{3}
C_{nc+}(i)=\sum _{j\in \Lambda_i} C_{nc}(j).
\end{equation}

Chen \emph{et~al.}  proposed a semi-local centrality measure as a tradeoff between low-relevant degree centrality
and other time-consuming measures (labeled as SL index). It considers both the nearest and the next nearest neighbors. The semi-local
centrality $SL(i)$ of node $i$ is defined as~\cite{chen2012identifying}
\begin{equation}\label{a}
Q(s)=\sum _{j\in \Lambda_s} N(j),
\end{equation}

\begin{equation}\label{b}
SL(i)=\sum _{s\in \Lambda_i} Q(s),
\end{equation}
where $\Lambda_i$ is the neighbor node set of node $i$. $N(j)$ is the number of the nearest and the next nearest
neighbors of node $j$.

It is fact that, on one hand, the influence of a node is increased if its neighbors (here the neighbors of a node do not just includes its nearest neighbors, which may also include next nearest neighbors, next-next nearest neighbors, etc.) have higher value of $ks$; on the other hand, the interaction effect between two nodes decreases with their distance. Enlighten by the idea of classical gravity formula proposed by Isaac Newton, we can view the k-shell value of node $i$ as its mass, and the shortest path distance between two nodes in network is viewed as their distance. In this way, the influence of node $i$ is measured by (labeled as G):
\begin{equation}\label{5}
G(i)=\sum_{j\in\psi_i}\frac{ks(i)ks(j)}{d_{ij}^2},
\end{equation}
where $d_{ij}$ is the shortest path distance between node $i$ and node $j$. $\psi_i$ is the neighborhood set whose distance to node $i$ is less than or equal to a given value $r$. To reduce the algorithm complexity, in the paper, we let $r=3$, i.e, only nearest neighbors, next nearest neighbors and the next-next nearest neighbors are considered.

An extended gravity index is further developed based on Eq.~(\ref{5}), which is defined as (labeled as $G_+$):
 \begin{equation}\label{6}
G_{+}(i)=\sum_{j\in\Lambda_i}G(j),
\end{equation}
 $\Lambda_i$ is the nearest neighborhood of node $i$.

\section{Experimental results} \label{sec:main results}

In this section, we compare the effectiveness of other indices with our $G$ or $G_+$ index from different aspects and on different networks, including real networks as well as synthetic networks.

We employ the standard susceptible-infected-removed (SIR) model~\cite{moreno2002epidemic} to estimate the \emph{real} spreading influence of the nodes (labeled by $R$). In detail, to check the spreading influence of one given node, we set this node as an infected node and the other nodes are susceptible nodes. At each time step, each infected node can infect its susceptible neighbors with infection probability $\beta$, and then it recovered from the diseases with probability $\mu$. In this paper, we set $\mu=1.0$. This process repeats until there has no any infected nodes. At last, the number of recovered nodes is used to reflect the \emph{real} influence of the node.  To guarantee the reliability of the results, all of them are at least averaged over 1000 independent realizations.

First, a small example network with $N=20$ nodes and $M=25$ edges is given in Fig.~1 to intuitively compare these indices, the ranking lists from different indices are presented in Table~1.
\begin{figure}
\begin{center}
\includegraphics[height=130mm,width=130mm]{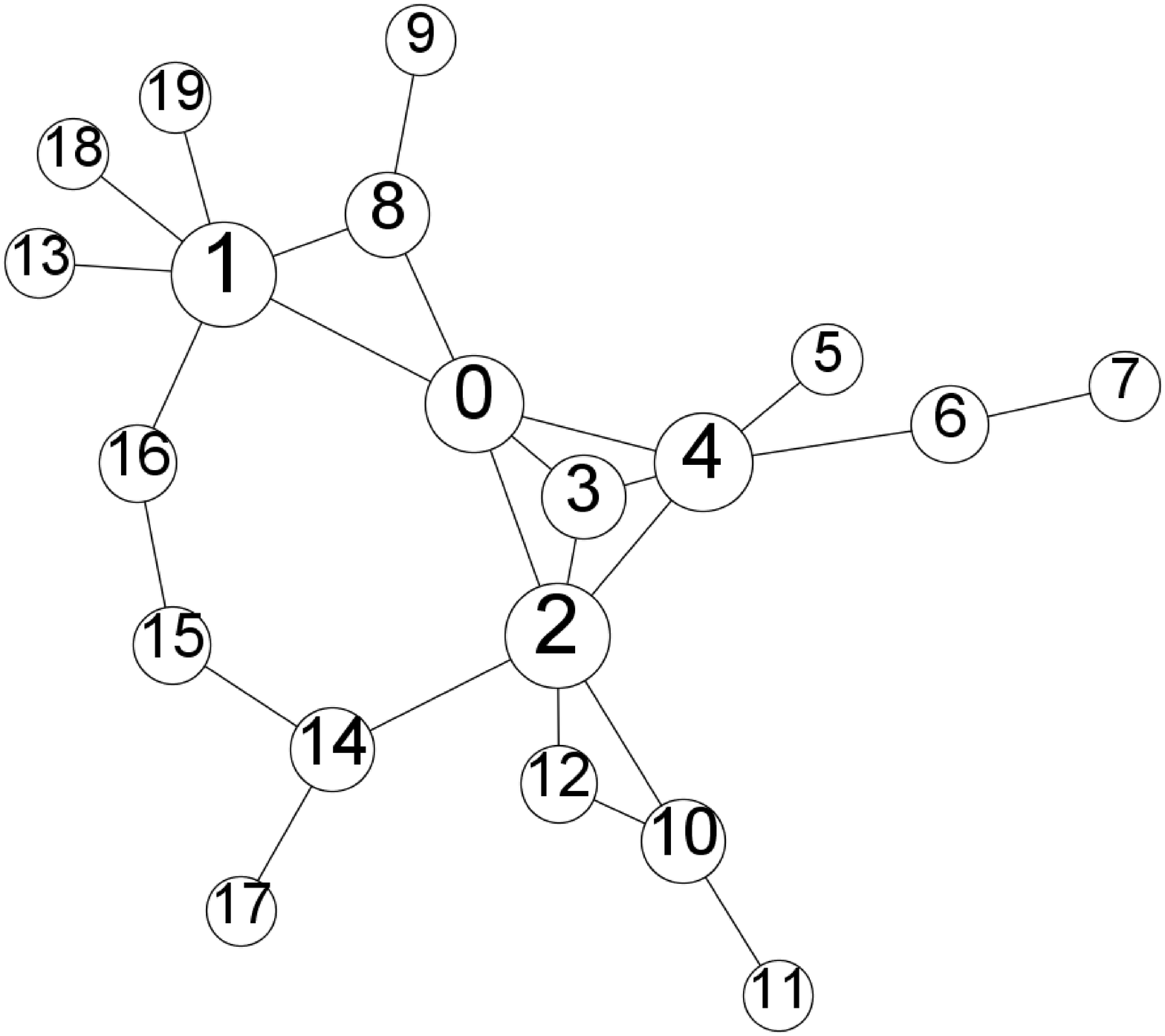}
\caption{ An example network with 20 nodes and 25 edges. Nodes with larger degrees have larger size. }\label{fig1}
\end{center}
\end{figure}
From Table~1, one can see that the $ks$ centrality cannot well distinguish the influence of nodes. Even in the same shell, the nodes' influence may be totally different. Moreover, the result indicates that, our proposed $G$ and $G_+$ index can effectively identify the influence of nodes, i.e., the ranking list determined from $G$ or $G_+$ index is in good agreement with the ranking list obtained from SIR model in the last row.

\begin{table}[h]\label{table1}
\scriptsize
\caption {The ranking lists determined by different indices. Degree centrality: DC; mixed Degree decomposition: MDD; gravity centrality: G; extended gravity centrality: $G_+$; extended neighborhood coreness defined in Eq.~(\ref{3}): $C_{nc+}$; k-shell decomposition: ks; betweeness centrality: BC; closeness centrality: CC; semi-local centrality measure: SL; the node spreading influence evaluated by SIR model: R, by taking $\beta=0.25$.}
\begin{tabular}{l|c|c|c|c|c|c|c|c|c|c}
\hline
Rank &DC &MDD &G &$G_+$ &$C_{nc+}$ &ks &BC &CC &SL &R\\
\hline
1 &1, 2 &1, 2 &2 &2 &0 &0, 2, 3, 4 &2 &0 &0 &2\\  \hline
2 &0,4 &0, 4 &0 &0 &2 &1, 8, 10, 12, 14-16 &0 &2 &2 &0\\ \hline
3 &3, 8, 10, 14 &3 &4 &4 &4 &others &1 &4 &4 &4\\ \hline
4 &6, 12, 15, 16 &8 &3 &3 &3 &--- &4 &1 &3 &1\\ \hline
5 &others &10,14 &1 &1 &1 &--- &14 &3 &1 &3\\ \hline
6 &--- &12, 15, 16 &8 &8 &8 &--- &6, 8, 10 &8 &8 &8\\ \hline
7 &--- &6 &14 &--- &10 &--- &16 &14 &10 &14\\ \hline
8 &--- &others &10 &--- &12, 14 &--- &15 &10 &--- &10\\ \hline
9 &--- &--- &--- &--- &--- &--- &others &--- &--- &---\\ \hline
\end{tabular}
\end{table}

To validate the effectiveness of the $G$ or $G_+$ index, we apply it to 9 real networks, including Facebook (Slavo Zitnik's friendship network in Facebook)~\cite{blagus2012self}, Netsci (collaboration network of network scientists)~\cite{newman2006finding}, Email (e-mail network of University at Rovira i Virgili, URV)~\cite{guimera2003self}, TAP(yeast protein-protein binding network generated by tandem affinity purification experiments)~\cite{zeng2013ranking}, Y2H (yeast protein-protein binding network generated using yeast two hybridization)~\cite{kumar2002proteomics}, Blogs (the communication relationships between owners of blogs on the MSN (Windows Live) Spaces website)~\cite{xie2006social}, Router (the router-level topology of the Internet)~\cite{spring2004measuring}, HEP (collaboration network of high-energy physicists)~\cite{leskovec2007graph}, PGP (an encrypted communication
network)~\cite{boguna2004models}. For simplicity, these networks
are treated as undirected and unweighed networks in this work. The detailed information about these 9 real networks are presented in Table~2.

\begin{table}\label{table2}
\centering
\caption {Basic structural properties. $N$ and $M$ are the number of nodes and edges, respectively. $\beta_{th}$ is the epidemic threshold. $H$ is degree heterogeneity, given by $\langle k^2\rangle/\langle k\rangle^2$. $\tilde{r}$ is assortativity coefficient. $C$ is clustering coefficient. $L$ is average shortest path length. $D$ is diameter.}
\begin{tabular}{l|c|c|c|c|c|c|c|c}
\hline
Network &N &M &$\beta_{th}$ &H &$\tilde{r}$ &C &L &D\\
\hline
Facebook &324 &2218 &0.047 &1.567 &0.247 &0.465 &3.054 &7\\ \hline
Netsci &379 &914 &0.125 &1.663 &-0.082 &0.741 &6.042 &17\\ \hline
Email &1133 &5451 &0.053 &1.942 &0.078 &0.220 &3.606 &8\\ \hline
TAP &1373 &6833 &0.061 &1.644 &0.579 &0.529 &5.224 &12\\  \hline
Y2H &1458 &1948 &0.140 &2.667 &-0.209 &0.071 &6.812 &19\\ \hline
Blogs &3982 &6803 &0.072 &4.038 &-0.133 &0.284 &6.252 &8\\ \hline
Router &5022 &6258 &0.072 &5.503 &-0.138 &0.012 &6.449 &15\\ \hline
HEP &5835 &13815 &0.110 &1.926 &0.185 &0.506 &7.026 &19\\ \hline
PGP &10680 &24316 &0.053 &4.147 &0.238 &0.266 &7.463 &24\\ \hline
\end{tabular}
\end{table}

How to improve the resolution is the key issue of an algorithm, for instance, in Ref.~\cite{zhou2009predicting}, Zhou \emph{et~al.} have clarified that the resolution problem is a major reason for the poor performance of common neighbor index subject to the AUC value in link predication, and then they proposed a local path index to solve the resolution problem. Similarly, a good index in ranking the influences of nodes should also has a high resolution. As illustrated in Table 1, $G$ or $G_+$ index is good at distinguishing the nodes' difference, which is much better than the $ks$ index. So to quantitatively measure resolution of different indices, a monotonicity index $M(X)$ for a ranking list $X$ is used~\cite{bae2014identifying}:
\begin{equation}\label{7}
M(X)=[1-\frac{\sum _{c\in V} N_{c}(N_{c}-1)}{N(N-1)}]^2,
\end{equation}
where $N$ is the size of network, and $N_{c}$ is the number of nodes with the same index value $c$.  If $M(X)=1$, which means that the ranking method is perfectly monotonic and each node is categorized a different index value; otherwise, all nodes are in the same rank as $M(X)=0$. The monotonicity $M$ for different ranking methods is summarized in Table~3. Generally, the results suggest that $G$ or $G_{+}$ index can give higher value of $M$. Moreover, $M(G)$ and $M(G_+)$ are very near 1 in some networks. Therefore, gravity method can better distinguish the node's influence than other indices.

\begin{table}[htbp]
\centering
\footnotesize
\caption {M (.) is the monotonicity of the corresponding measures.}
\scriptsize
\begin{tabular}{l|c|c|c|c|c|c|c|c|c}
\hline
Network &M(DC) &M(MDD) &M(G) &M(G+) &M(Cnc+) &M(ks) &M(BC) &M(CC) &M(SL)\\
\hline
Facebook &0.9315 &0.9729 &0.9999 &0.9995 &0.9995 &0.8445 &0.9855 &0.9953 &0.9999\\ \hline
Netsci &0.7642 &0.8215 &0.9949 &0.9951 &0.9893 &0.6421 &0.3387 &0.9928 &0.9939\\ \hline
Email &0.8874 &0.9229 &0.9999 &0.9999 &0.9991 &0.8088 &0.9400 &0.9988 &0.9999\\ \hline
TAP &0.8991 &0.9599 &0.9994 &0.9994 &0.9981 &0.8380 &0.9238 &0.9988 &0.9992\\  \hline
Y2H &0.4884 &0.5304 &0.9966 &0.9960 &0.9633 &0.2972 &0.5063 &0.9957 &0.9936\\ \hline
Blogs &0.5654 &0.5906 &0.9976 &0.9976 &0.9868 &0.4670 &0.4004 &0.9973 &0.9971\\ \hline
Router &0.2886 &0.3009 &0.9967 &0.9965 &0.9657 &0.0691 &0.2983 &0.9961 &0.9953\\ \hline
HEP &0.7654 &0.8314 &0.9998 &0.9999 &0.9917 &0.6303 &0.5651 &0.9998 &0.9990\\ \hline
PGP &0.6193 &0.6678 &0.9995 &0.9997 &0.9851 &0.4806 &0.5099 &0.9996 &0.9986\\ \hline
\end{tabular}
\end{table}
\normalsize

The Kendall's tau rank correlation coefficient $\tau$ is used to measure the correlation one topology-based ranking list and the real spreading capability $R$. Let $(x_{i},y_{i})$ and $(x_{j},y_{j})$ be a randomly selected pair of joint observations from ranking lists $X$ and $Y$, respectively. If one has $x_{i}>x_{j}$ and $y_{i}>y_{j}$ or $x_{i}<x_{j}$ and $y_{i}<y_{j}$, the observations $(x_{i},y_{i})$ and $(x_{j},y_{j})$ are said to be concordant. If $x_{i}>x_{j}$ and $y_{i}<y_{j}$ or $x_{i}<x_{j}$ and $y_{i}>y_{j}$, they are said to be discordant. If $x_{i}=x_{j}$ or $y_{i}=y_{j}$, the pair is neither concordant nor discordant~\cite{knight1966computer,zhou2009predicting}. $\tau$ is defined as
\begin{equation}\label{system}
\tau=\frac{N_{1}-N_{2}}{0.5N(N-1)},
\end{equation}
where $N_{1}$ and $N_2$ are the number of concordant pairs and discordant pairs, respectively.

When we employ SIR model to check the spreading influence of nodes, the infection probability $\beta$ should not be too small or too large. The epidemic cannot successfully spread over networks if $\beta$ is too small, so the spreading capability of each node cannot be measured. On the contrary, if $\beta$ is too large, the epidemic can easily outbreak over almost whole network, leading to the spreading capability of each node cannot be distinguished too. Thus, in this work, we first obtain the epidemic threshold $\beta_{th}$ for each network, which is given as $\beta_{th}\sim\langle k\rangle/\langle k^2\rangle$, with $\langle k\rangle$ and $\langle k^2\rangle$ be the average degree and the second order average degree~\cite{moreno2002epidemic}, respectively. The value of $\beta_{th}$ for different networks is given in Table 2 too.  Then, we choose the value of $\beta$ to be slightly larger than the threshold $\beta_{th}$ when computing $\tau$ for different indices (a new index to measure the influence of nodes was proposed in Ref.~\cite{lin2015locating}, which is independent on the parameter $\beta$). The results in Table~4 manifest that our method outperforms the other methods in most cases.

\begin{table}[htbp]\label{table3}
\centering
\footnotesize\caption {$\tau(.)$ is correlation of corresponding methods for given $\beta$.}
\scriptsize
\begin{tabular}{l|c|c|c|c|c|c|c|c|c|c}
\hline
Network  &$\beta$ &$\tau_{DC}$ &$\tau_{MDD}$ &$\tau_{G}$ &$\tau_{G_+}$ &$\tau_{C_{nc+}}$ &$\tau_{ks}$ &$\tau_{BC}$ &$\tau_{CC}$ &$\tau_{SL}$\\
\hline
Facebook &0.050 &0.767 &0.796 &0.861 &0.913 &0.916 &0.735 &0.364 &0.720 &0.940\\  \hline
Netsci &0.130 &0.599 &0.620 &0.830 &0.852 &0.847 &0.525 &0.308 &0.330 &0.806\\ \hline
Email &0.070 &0.771 &0.790 &0.887 &0.937 &0.935 &0.779 &0.625 &0.822 &0.935\\ \hline
TAP &0.065 &0.725 &0.746 &0.870 &0.899 &0.873 &0.690 &0.273 &0.527 &0.886\\ \hline
Y2H &0.160 &0.445 &0.463 &0.827 &0.833 &0.825 &0.407 &0.412 &0.701 &0.775\\ \hline
Blogs &0.075 &0.525 &0.532 &0.834 &0.763 &0.795 &0.482 &0.390 &0.579 &0.706\\ \hline
Router &0.075 &0.322 &0.323 &0.797 &0.805 &0.786 &0.186 &0.315 &0.642 &0.790\\ \hline
HEP &0.110 &0.487 &0.506 &0.787 &0.865 &0.735 &0.485 &0.345 &0.784 &0.840\\ \hline
PGP &0.055 &0.479 &0.490 &0.784 &0.770 &0.756 &0.439 &0.313 &0.636 &0.747\\ \hline
\end{tabular}
\end{table}

To further estimate how the infection probability $\beta$ affects the effectiveness of different methods, the correlation value $\tau$ as a function of $\beta$ for different methods is shown in Fig.~2. As described in Fig.~2, in most cases, $G$ or $G_+$ index provides better performance than the other index when $\beta>\beta_{th}$ (the values of $\beta_{th}$ for different networks are illustrated by the dot lines in Fig.~2). However, Fig.~2 clearly indicates that though the global indices, such as betweenness index and closeness index  are time-consuming, they are not good at measuring the influence of nodes in these networks. Meanwhile, the performance of MDD method in identifying the node's influence is almost the same as the degree centrality.

Previously, our results were obtained by setting the value of $r=3$ (i.e., only the effects of the nearest neighbors, next nearest neighbors and the next-next nearest neighbors are considered). To check the sensitivity of $r$ on our results, the effect of the parameter $r$ on the value of $\tau$ is plotted in Fig.~3. As shown in Fig.~3, in generally, the optimal value of $r$ is about 3-5, and the value of $\tau$ becomes stable when $r$ is further increased. As a result, it is unnecessary to choose too large value of $r$, which just increases the algorithm complexity of our method.

\begin{figure}
\begin{center}
\includegraphics[height=100mm,width=150mm]{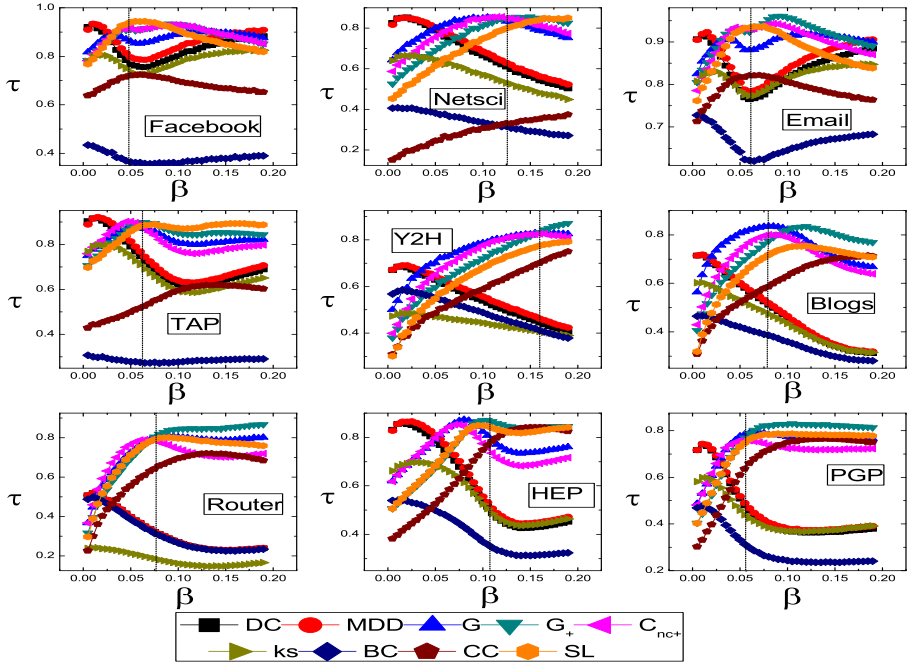}
\caption{(Color online)  The value of $\tau$ obtained by comparing the ranking list generated by the SIR model and the ranking lists generated by the topology-based method on Facebook, Netsci, Email, TAP, Y2H, Blogs, Router and HEP. The dot lines correspond to the epidemic threshold. }\label{fig2}
\end{center}
\end{figure}

\begin{figure}
\begin{center}
\includegraphics[height=100mm,width=130mm]{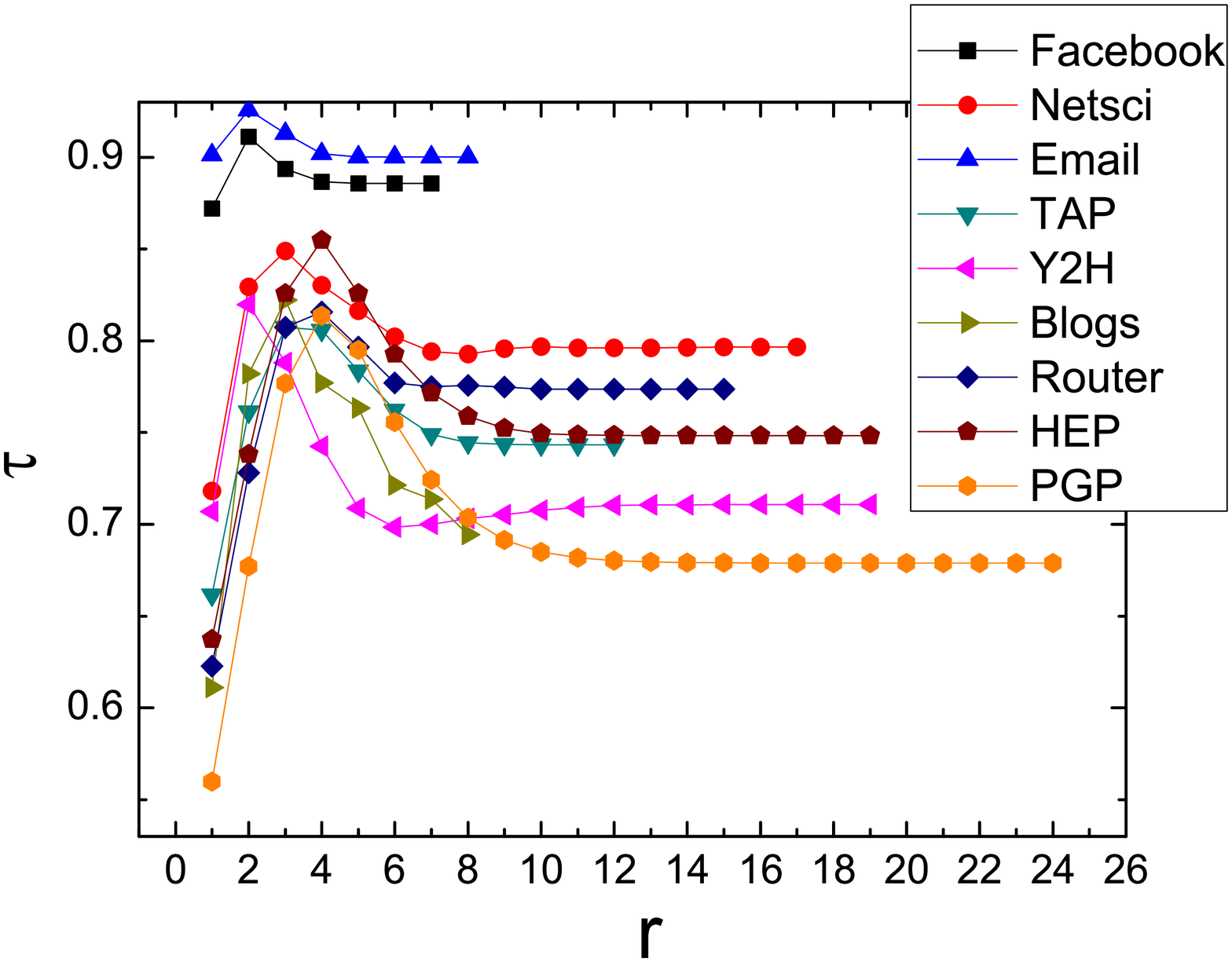}
\caption{(Color online) The effect of considered range r on the Kendall's tau rank correlation coefficient $\tau$. Here the value of $\beta$ for each network is the same to the value of $\beta$ in Table 4. We should address that, though the average distance of Facebook is about 3, the value of $r$ is larger than 3, since the distance between a pair of randomly chosen nodes may be larger than 3. Therefore, the largest value of $r$ is the diameter of the network. For example, $r=7$ for the Facebook network.}\label{fig3}
\end{center}
\end{figure}

Besides the real networks, we also compare the performance of our method with other methods on the two typical synthetic networks--Barab\'as-Albert (BA) networks~\cite{barabasi1999emergence} and the Watts-Strogatz (WS) small-world networks~\cite{watts1998collective} with $N=1000$.  Starting from a connected network with $m_{0}$ nodes to construct a BA network, at each step, a new node is added to the network and connected to $m$ existing nodes according to the preferential attachment mechanism, where $m\leq m_{0}$~\cite{barabasi1999emergence}. We set the number of nodes $m_{0}=10$ in this paper.
The WS small-world model considers a ring nearest neighbor coupled network with $N$ nodes. Each node symmetrically connects to its $2K$ nearest neighbors. Starting from it, a fraction $p$ of edges in the
network are rewired, by visiting all $K$ clock-wise edges
of each node and reconnecting them, with probability
$p$, to a randomly chosen node~\cite{watts1998collective}. During the rewiring process, self-connection and reconnection are forbidden.

For BA network (see Fig.~4 (a) and (b)), one can see that the performances of $G$, $G_+$ and $C_{nc+}$ indices are almost the same. The reason is that the three indices are all the improved methods of k-shell method, however, all nodes in BA network are almost classified into the same shell when using the the k-shell method (so we do not calculate the case of $ks$ in Fig.~4). Moreover, the results show that the three indices are better than CC index and are  much better than DC, BC and MDD indices. For WS network (see Fig.~4 (c) and (d)), whose degree distribution shows Poisson distribution, i.e, their degrees are not so different. In this case, it is difficult for DC index to distinguish the influence of nodes. However, as shown in Fig.~4(c) and (d), as $\beta>\beta_{th}$, the performances of $G$ and $G_+$ indices are still better than the other indices. In particular, for WS network, one can observe that the performances of $G$ and $G_+$ indices are much better than the $C_{nc+}$ index when $\beta>\beta_{th}$. The results in Fig.~4 suggest that our method can not only identify the influential nodes on real networks but also on synthetic networks.
\begin{figure}
\begin{center}
\includegraphics[height=100mm,width=130mm]{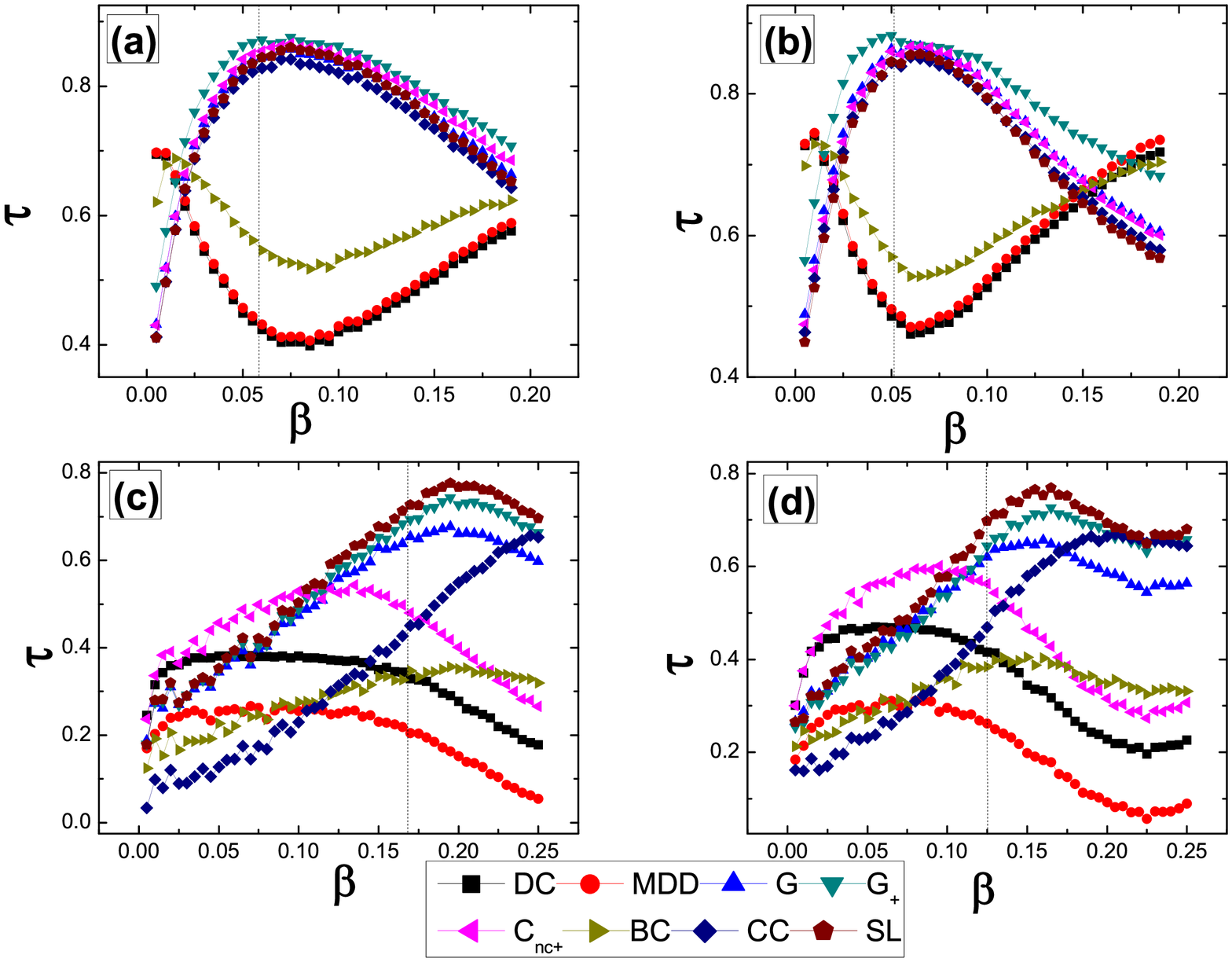}
\caption{(Color online) (a) BA: $m=3$; (b) BA: $m=4$;(c) WS: $K=3$, $p=0.05$; (d) WS: $K=4$, $p=0.05$. The pink dot lines correspond to the epidemic threshold.}\label{fig3}
\end{center}
\end{figure}

\section{Conclusions and discussions} \label{sec:discussion}
In summary, in this paper, we have proposed a gravity method to identify the influential spreaders in complex networks. In the model, each node's k-shell value is considered as its mass and the shortest path distance between two nodes is viewed as their distance. The idea of the gravity method comes from the well-known gravity formula, which is very dramatic and impressive. What's more, the gravity model can reflect the facts that, on one hand, the interaction influence between two nodes is proportional to their corresponding k-shell values; on the other hand, the influences of the neighbors decreases with their distance. We employed our method
on some real networks and synthetic networks, by calculating the monotonicity index $M$, we found that our method can better distinguish the difference of node influence than other indices. Also, by computing Kendall¡¯s tau rank correlation coefficient $\tau$, we have shown that, in most cases, our method has a better performance in evaluating the node's influence than other indices. Therefore, our method provides an effective way to identify the influential spreaders in social networks.

Some extensions may be made based on this method. For example, by defining the combination of node's degree and node's strength as the weighted degree of a node in weighted networks, Garas \emph{et~al} have proposed a new k-shell decomposition method for weighted networks~\cite{garas2012k}. Therefore, once the new k-shell value for each node in weighted network is assigned, our method can be simply generalized to weighted networks~\cite{li2014identifying}. Also, if we view the closeness centrality, degree centrality, eigenvector centrality, and so forth as the mass of a node, then the gravity method may be further generalized.

We only investigated the performance of the gravity method in some typical networks, and the classical SIR model was used to mimic the the spreading dynamics. In reality, the structure of networks and the spreading dynamics are diverse. For example, recent researches have illustrated that networks in nature do not act in isolation, but instead exchange information and depend on one another to function properly, that is to say, natural systems are organized in interconnected networks~\cite{buldyrev2010catastrophic,reis2014avoiding,gallos2012small}; And some real spreading dynamics like the diffusion of rumors or opinions, the rise of scientific ideas~\cite{pei2014searching,pei2013spreading,hu2014conditions} are different from the spreading of epidemic, which may challenge the effectiveness of proposed indices. For example, Borge-Holthoefer \emph{et~al.} have stated that the influential nodes in networks are absent when considering the rumor dynamics~\cite{borge2012absence}, also in Ref~\cite{klemm2012measure}, authors have illustrated that the roles of nodes are dependent on the collective dynamics. Therefore, feasible methods need to be examined, we here hope our work inspire possible solutions to the above mentioned problems in the near future.

\section*{Acknowledgments}

This work is funded by the NSFC
(Grant Nos. 61473001, 11331009, 11275186). Partially supported by open fund of Key Laboratory of Computer Network and Information Integration (Southeast University ), Ministry of Education (No. K93-9-2015-03B), and by Graduate Student Academic Innovation Foundation of Anhui University (Grant No. yfc100015).

\end{document}